\documentclass[twocolumn,showpacs,preprintnumbers,amsmath,amssymb]{revtex4}

\usepackage{graphicx}
\usepackage{dcolumn}
\usepackage{bm}

\begin{document}


\title{Ranking by Loops: a new approach to categorization}

\author{Valery Van Kerrebroeck and Enzo Marinari}
\affiliation{%
\normalsize{Dipartimento di Fisica, ``Sapienza'' Universit\`a di Roma,}\\
\normalsize{INFM-CNR and INFN, P.le A. Moro 2, 00185 Roma, Italy.}\\
}%

\date{\today}

\begin{abstract}
We introduce Loop Ranking, a new ranking measure based on the detection 
of closed paths, which can be computed in an efficient way. 
We analyze it with respect to several ranking measures which have been 
proposed in the past, and are widely used to capture the relative 
importance of the vertices in complex networks. 
We argue that Loop Ranking is a very appropriate measure to 
quantify the role of both vertices and edges in the network traffic.
\end{abstract}

\maketitle

Finding the most important vertices is an important problem in complex
network analysis \cite{internet-1}.  In technological
networks, such as the Internet, the main hubs (i.e., the vertices with
many links to other vertices) play an important role in the stability
of the network \cite{internet}.  Instead, the removal 
of any kind of species in food webs may cause the disintegration of 
the corresponding network \cite{foodweb}.  
The order of importance
of the vertices is referred to as a \emph{ranking}.  In general the
importance of a vertex, and thus its ranking, depends very much on the
type of network which is under consideration.  The measures that are
based on topological features are assumed to be more objective, and
typically result in more adequate rankings.

The most simple topology based ranking measure is the
degree, i.e., the number of neighboring vertices of a vertex.  While
the degree is easy to compute, it is not a very refined measure, as it
solely depends on the local neighborhood around a vertex.  Several
more global measures ~\cite{eigentrust, hits, pagerank-theory, bickson, centrality} have
been proposed which do take the overall structure of the network into
account.  Apart from finding a fast way to compute these measures,
another challenge is usually to decide which of these measures are
more appropriate for which type of network.

We start our reasoning from the observation that the link structure of
complex networks describes the topology of interactions taking place
in a dynamic complex system: the importance of a vertex (or
edge) to the network traffic is related to the number of paths or
walks it lies on \cite{Vattay}.

The topology of most real-world networks is represented by a directed
graph $G(V,E)$ which is completely characterized by its set $V$ of $N$
vertices $i$ and by the set $E$ of $M$ directed edges $i \to j$, where
$i$ and $j$ are said to be the starting and ending vertex,
respectively, of the directed edge $i \to j$.  For undirected graphs
either one of the ending vertices $i$ and $j$ of the edge $\{i,j\}$
can be the ending or starting vertices.  Accordingly, information can
travel in either direction along the edges of an undirected network.
In case of weighted directed graphs, we associate a weight
$r_{i\to j}$ to each one of the directed edges $i \to j$.  For
unweighted graphs all edge weights are uniformly equal to one.  We
denote the set of edges or the set of vertices of these edges distinct
from $i$ (as it is always clear from the context) of which $i$ is an
ending vertex or starting vertex as $\partial^{+}_i$ or
$\partial^{-}_i$, respectively.  Correspondingly, the in-degree
$d^{+}_i$ and out-degree $d^{-}_i$ of a vertex $i$ are defined as the
sum of the weights of the edges of which that vertex $i$ is an ending
vertex, or starting vertex, respectively.

Walks are defined as sequences of vertices $(i_1,\ldots,i_L)$,
where for each couple of subsequent vertices $i_{k-1}$ and $i_{k}$,
for $k=2,\ldots,L$, the directed edge $i_{k-1} \to i_k$ belongs to
$E$.  As such, they can cross the same edge or vertex infinitely many
times.  Instead, paths are defined as self-avoiding walks.  In
particular, a cycle or loop is a closed path. More formally, it is
defined as a sequence $(i_1,i_2\ldots,i_L,i_1)$ of vertices, where for
all $k=1,2,\ldots,L$ these $i_k$ are distinct from each other, and,
for all $k=2,\ldots,L$, each couple of subsequent vertices $i_{k-1}$
and $i_{k}$ are connected by a directed edge $i_{k-1} \to i_{k}$
belonging to $E$, as does $i_L \to i_1$.
The weight of any of these subgraphs is defined as the product of 
the weights $r_{i \to j}$ of the edges composing that subgraph. 
In case of a cycle defined by a sequence $(i_1,i_2,\ldots,i_L,i_1)$, 
its corresponding weight is given by  
$w[\rm C]=r_{i_L \to i_1}\prod_{k=2}^{L}r_{i_{k-1} \to i_{k}}$.

One measure which is widely used to find the most relevant vertices of
a network is PageRank \cite{pagerank-theory}.  PageRank is an
iteratively computed ranking measure where the PageRank of a given
vertex depends on the PageRank of its neighboring vertices.  More
formally, the PageRank ${\cal P}(i)$ of a vertex $i$ is defined as
\begin{equation}
{\cal P}(i)=c \sum_{j \in {\partial^{+}_i} } 
\frac{{\cal P}(j)}{d^{-}_{j}} +\frac{1-c}{N}\;,
\label{eq:pagerank}
\end{equation}
where $c$ is a damping factor chosen in the interval $]0,1]$.  In
matrix form this becomes ${\cal P}=c\; {\rm C}^T {\cal P}
+\frac{1-c}{N}\;{\delta}$, where 
$\cal P$ is a $N$-dimensional
vector, $\delta$ is a $N$-dimensional
vector with all elements equal to one,
and the elements $C_{ij}$ of the $N\times N$ matrix ${\rm C}$
are equal to $1/d^{-}_{i}$ if the edge $i \to j$ belongs to $E$, 
and zero otherwise.
As the sum of the
entries of a column of this matrix ${\rm
  C}$ is equal to one, it can be interpreted as a Markov matrix.  The
resulting PageRank is then proportional to the probability with which
a random walker will come across a given vertex.  As such, PageRank is
an importance measure for vertices which, being based on random walks,
takes the overall structure of the network into account.  However, the
question naturally arises whether it is not preferable to emulate the
behavior of a more efficient \emph{self-avoiding} random walker.

Motivated by the latter observation, we introduce a new ranking
measure based on paths rather than walks.  Several centrality measures
based on either the number or length of shortest paths passing through
or ending at a given vertex have already been proposed
\cite{centrality}.  In particular, the Betweenness Centrality ${\cal
  B}$ (BC) of a given vertex (or edge) is defined as the fraction of
shortest paths on which that vertex (or edge) lies. Defining
$\sigma_{k,l}$ as the number of shortest paths between the vertices
$k$ and $l$, and $\sigma_{k,l}(i)$ and $\sigma_{k,l}(i \to j)$ the
number of these passing through vertex $i$ or edge $i \to j$, we have
\begin{equation}
{\cal B}(i)= \sum_{k,l (\neq i) \in V}\frac{\sigma_{k,l}(i)}{\sigma_{k,l}}
\mbox{ and }
{\cal B}(i \to j)= \sum_{k,l \in V}\frac{\sigma_{k,l}(i \to j)}{\sigma_{k,l}},
\label{eq:betweenness_centrality}
\end{equation}
for the BC of vertex $i$ and the edge $i \to j$,
respectively.  One fundamental problem that prevents these measures
from becoming widely used in real network analysis is that they cannot
be computed as fast as, for example, PageRank \cite{centrality}.
Moreover, a measure based on shortest paths only may not be adequate
enough as also longer paths (with possibly higher weights) could add
to the centrality of a given vertex, or edge \cite{Vattay}.

We propose a ranking based on the presence of closed paths, i.e.,
cycles, through a given vertex.  We consider the probability of
presence of cycles, rather than all, i.e., also open, paths, for a
specific reason. Namely, it allows us to compute the corresponding
ranking by means of Belief Propagation, a distributed, message passing
algorithm which converges in linear time in the system size to the
marginal probabilities of presence of these
cycles.  This restriction to closed paths results in a ranking which
reflects the geometric position of each one of the vertices, and gives
a subjective view of how each vertex sees the overall network based on
paths.

In particular, we propose a ranking based on the sums of the weights of
these cycles. As such it represents the probability with which a
self-avoiding walker returns to the same vertex while exploring the
network, taking the weight of each path into account.  We define the
Loop Ranking ${\cal L}$ of a vertex $i$, or edge $i \to j$, 
as the sum of the weights $w$ of all cycles ${\rm C}$ passing 
through that vertex $i$, or edge $i \to j$ respectively, i.e.,
\begin{equation}
{\cal L}(i)= \sum_{{\rm C}\ni i} w[{\rm C}]
\mbox{ and }
{\cal L}(i \to j)= \sum_{{\rm C}\ni (i \to j)} w[{\rm C}]\;.
\label{eq:cycle-sum}
\end{equation} 
In practice, we do not compute the actual Loop Ranking, but rather the
marginal expressing the probability with which a cycle passes through
a given vertex, i.e., ${\cal L}(i)/\sum_{\rm C} w[{\rm C}]$, which
produces the same ordering.  The latter can be obtained by
reformulating the problem of identifying all cycles of a given graph
as a constraint satisfaction problem \cite{MaSe, MaSeVVK, MaVVK}.

We define an appropriate phase space in which all (simple) subgraphs,
such as cycles, are represented by a unique configuration.  To this
purpose we associate with each edge $(i \to j)$ an Ising-like variable
$S_{i\to j}$, where $S_{i\to j}$ takes on the value zero or one if the
corresponding edge $(i \to j)$ belongs, or does not belong to the
considered subgraph, respectively.  In this way, we establish
the desired one-to-one correspondence between any simple subgraph of
the original graph ${\rm G}$ and all configurations defined by any of
$2^M$ sequences $\underline{S}=(S_1,\ldots,S_M)$.  For simplicity, we
also introduce the notation $\underline{S}_i$ which denotes the set of
all edge variables $S_{i\to j}$ and $S_{j \to i}$ of which $i$ is a
starting or ending vertex, respectively.

We can now define the probability law 
\begin{equation}
{\rm Prob}[\underline{S}] = \frac{1}{Z} \prod_{(i \to j)\in E}(r_{i\to j})^{S_{i\to  j}} \prod_{i \in V} f_i(\underline{S}_i)\;,	
\label{eq:prob_UC}
\end{equation}
where the local constraints $f_i (\underline{S}_i)$ are equal to one
if $\sum_{j \in \partial^{+}_i}S_{ij}=\sum_{j \in \partial^{-}_i}S_{ij}$ is zero or one, 
while $f_i(\underline{S}_i)$ are equal to zero otherwise, and $Z$ is a
normalization constant.  The complete set of local constraints $f_i$
ensures that only those configurations representing subgraphs composed of
(possibly vertex disjoint) directed cycles have a non zero probability
(\ref{eq:prob_UC}).  The first product appearing in relation
(\ref{eq:prob_UC}) makes the probability of the allowed configurations
proportional to the weight of the subgraph they represent.

An approximation to the marginals of (\ref{eq:prob_UC}) can be
obtained using a local Monte Carlo like algorithm as presented in
\cite{MaSeVVK}.  However, for factorizable probability
laws, such as (\ref{eq:prob_UC}), they can also easily be computed
by means of message passing algorithms, such as Belief and Survey 
Propagation \cite{survey}. 
Belief Propagation (BP) is a distributed, iterative algorithm which is
intrinsically linear in the system size.  It requires the introduction
of $2M$ real-valued message variables, $M$ of type $x_{i \to j}$ 
in the same direction of the edges $S_{i \to j}$, and $M$ of type $y_{j
  \to i}$ going in the opposite direction.  Initially, 
they all take on a random value
in the interval $[0,1]$.  Each BP iteration then consists in an update
of these $2M$ variables.  Assuming (\ref{eq:prob_UC}), the update
rules have the following form,
\begin{equation}
x_{i \to j}  =
\frac{\sum_{k \in \partial^{+}_i} r_{k \to i} x_{k \to i}}
{ 1+ \underset{k \in \partial^{+}_i}{\sum} r_{k \to i} x_{k \to i}
\underset{k' \in \partial^{-}_i\setminus j}{\sum} r_{i \to k'} y_{k' \to i}} ,
\nonumber\label{eq:update-rule-1}
\end{equation}
\begin{equation}
y_{j \to i}  =
\frac{\sum_{k \in \partial^{-}_j} r_{j \to k} y_{k \to j}}
{ 1+ \underset{k \in \partial^{+}_j \setminus i}{\sum} r_{k \to j} x_{k \to j}
\underset{k' \in \partial^{-}_j}{\sum} r_{j \to k'} y_{k' \to j}} \;.
\nonumber\label{eq:update-rule-2}
\end{equation}
On acyclic graphs, the successive repetition of these BP iteration
steps always leads to a fixed point solution.  For generic graphs
containing cycles, BP does not necessarily converge \cite{bethe-BP}.
However, at least for sparse graphs, which do not contain too many
small loops and locally are tree-like, usually it does.

Once the fixed point has been reached, the corresponding value of the
message variables can be used to obtain the desired marginal
probabilities.  In our particular case, we are interested in the
vertex and edge marginals expressing the probability with which a
cycle contains that particular vertex or edge. Upon convergence of the
BP algorithm, these marginals can be obtained from the message
variables at the fixed point as
\begin{eqnarray}
\nonumber
&p_i&  = \frac{\sum_{k \in \partial^{+}_i} r_{k \to i} x_{k \to i} \sum_{k' \in \partial^{-}_i} r_{i \to k'} y_{k' \to i}}{1+\sum_{k \in \partial^{+}_i} r_{k \to i} x_{k \to i} \sum_{k' \in \partial^{-}_i} r_{i \to k'} y_{k' \to i}}\;, \\
\nonumber
\mbox{and }
&p_{i \to j}&  = \frac{r_{i \to j}^2 x_{i \to j}y_{j \to i}}{1+r_{i \to j}^2 x_{i \to j}y_{j \to i}}\;,
\end{eqnarray}
respectively.  On a generic, cyclic graph, the above expressions are
an approximation to the actual vertex and edge marginals. However, in
general, these approximations are very reasonable to work with.  The
marginals of (\ref{eq:prob_UC}) actually express
the probability with which a vertex, or edge, is part of a subgraph
composed of possibly several vertex disjoint directed cycles. Thus,
exact marginals of (\ref{eq:prob_UC}) are possibly an overestimation
of the desired marginals expressing the probability of presence of a
(single) closed path. However, for weighted graphs where the edge
weights have been rescaled such that they all lie in the interval
$[0,1]$, this effect is largely reduced, and has only a minor impact
on the resulting ordering \cite{MaVVK}.

\begin{figure}[!t]
\centering
\includegraphics[scale=0.75]{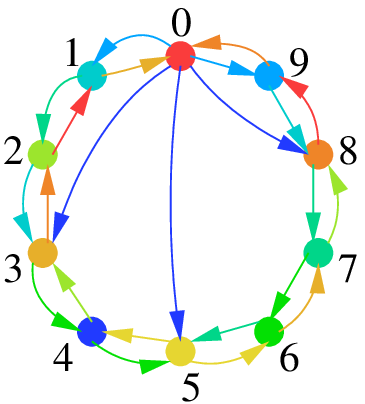}
\hspace{0.5cm}
\includegraphics[scale=0.75]{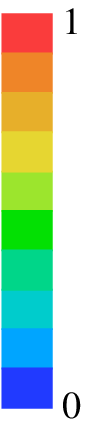}
\hspace{0.5cm}
\includegraphics[scale=0.75]{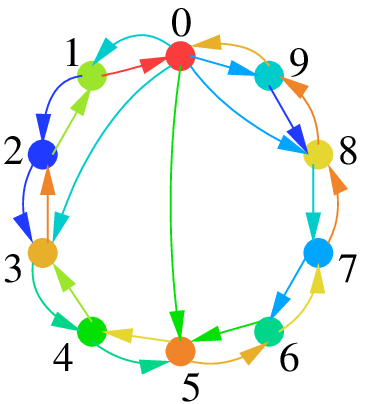}
\caption{\label{fig:vertexedgeranking_Gout1} 
Directed small world network $G_1$. 
The various colors express the ranking of the vertices and edges 
based on Loop Ranking (on the left) and Betweenness Centrality 
(on the right).}
\end{figure}

In order to get hints about which is a good ranking measure for
dynamic networks, we have looked at a number of examples and
particular cases.  We discuss here explicitly the case of two directed
small world networks that we consider very telling.  We assume them to
be unweighted for simplicity.  We first consider the graph shown in
figure \ref{fig:vertexedgeranking_Gout1}.  If we only take the outer
ring of edges into account, all information can be exchanged between
any two vertices in two ways, either in clockwise or counter-clockwise
direction, in which case all vertices and edges are considered to be
equally important.

The presence of the directed ``short cut'' edges (which are all
out-bound from vertex $0$) reduces the length of the shortest paths, a
feature which is typical of most real world networks \cite{networks}.
The resulting ``smaller world'' (from the point of view of a single
vertex) has a large impact on the mobility inside a network.  In
particular, the vertex $0$ will play a more crucial role than other
vertices in dispatching packages along the network.  Similarly, though
in a minor way, the ending vertices of the extra out-bound short cuts,
i.e., 3, 5 and 8, should play a more important role in the traffic
along this network.  Also, we expect the presence of these edges to
break the symmetry of the role of the single edges in the network
flow.

The results according to the various ranking schemes and corresponding
ordering for the vertices of graph $G_1$ are reported in table
\ref{tab:G_vertex}. We have rescaled the PageRank, Loop Ranking and
BC results such that they all lie in the interval 
$[0,1]$.  Clearly, the different ranking schemes attribute various
degrees of importance to the vertex $0$.  As the PageRank of a vertex
depends primarily on the number of edges directed towards that vertex
and the rescaled PageRank they transmit (rescaled by the out-degree of
their respective starting vertices), vertex $0$ of graph $G_1$ has a
relative low PageRank. Instead, as it does lie on most of the (closed)
paths of graph $G_1$, its central role is acknowledged by both the
Loop Ranking as the BC.  Moreover, the latter two
rankings recognize the increased role of the vertices 3, 5 and 8 with
respect to the other vertices, while PageRank makes no clear-cut
difference between them.
For the other vertices, the
ordering does differ depending on which path based ranking is
considered, as Loop Ranking depends on all closed paths, while
BC is only based on the number of shortest paths.

\begin{table}
\caption{\label{tab:G_vertex} Ranking of the vertices of the 
small world graph $G_1$ and $G_2$ shown in figure~\ref{fig:vertexedgeranking_Gout1} and~\ref{fig:vertexedgeranking_Gin1}, respectively,
according to the various ranking measures. The corresponding values of 
their PageRank, Loop Ranking and Betweenness Centrality, rescaled 
such that they lie in the interval $[0,1]$ 
(e.g., $({\cal L}(i)-{\cal L}_{\min})/({\cal L}_{\max}-{\cal L}_{\min})$), 
are also included.}
\begin{ruledtabular}
\begin{tabular}{ccccccccccccc}
\multicolumn{2}{c}{${\cal P}_{G_1}(i)$} & \multicolumn{2}{c}{${\cal L}_{G_1}(i)$} & \multicolumn{2}{c}{${\cal B}_{G_1}(i)$} & \multicolumn{2}{c}{${\cal P}_{G_2}(i)$} & \multicolumn{2}{c}{${\cal L}_{G_2}(i)$} & \multicolumn{2}{c}{${\cal B}_{G_2}(i)$} \\

\hline
5  &  1.00   &0  &  1.00 & 0  &  1.00 &  0  &  1.00    &  0  &  1.00      &  0  &  1.00  \\
4  &  0.88   &8  &  0.57 & 5  &  0.47 &  1  &  0.62     &  8  &  0.57       &  5  &  0.45  \\
3  &  0.83   &3  &  0.47 & 3  &  0.31 &  9  &  0.53     &  3  &  0.47       &  3  &  0.28  \\
6  &  0.81   &5  &  0.41 & 8  &  0.26 &  2  &  0.31     &  5  &  0.41       &  8  &  0.25  \\
7  &  0.68   &2  &  0.30 & 1  &  0.17 &  8  &  0.31     &  2  &  0.30       &  1  &  0.20  \\
8  &  0.58   &6  &  0.14 & 4  &  0.10 &  3  &  0.17     &  6  &  0.14       &  4  &  0.10  \\
2  &  0.43   &7  &  0.07 & 6  &  0.10 &  7  &  0.09    &  7  &  0.07      &  6  &  0.09  \\
0  &  0.08   &1  &  0.05 & 9  &  0.06 &  5  &  0.05    &  1  &  0.05      &  9  &  0.06  \\
9  &  0.07   &9  &  0.03 & 7  &  0.05 &  6  &  0.04     &  9  &  0.03      &  7  &  0.03  \\
1  &  0.00  &4  &  0.00 & 2  &  0.00 &  4  &  0.00   &  4  &  0.00      &  2  &  0.00  \\ 
\end{tabular}
\end{ruledtabular}
\end{table}

The importance ordering of both the vertices and the edges produced by 
Loop Ranking and 
BC is schematically presented by figure
\ref{fig:vertexedgeranking_Gout1}.
As in case of the vertices, Loop Ranking and
BC do not produce the exact same ordering of
edges, but there are no essential huge shifts between the two
corresponding rankings.
Note that edges with high ranking usually connect one vertex with high 
and another with low ranking.
The presence of these edges has been observed in case of protein networks
\cite{protein}, where it was argued that they play a crucial role in 
the overall robustness of the network.

\begin{figure}[!t]
\centering
\includegraphics[scale=0.75]{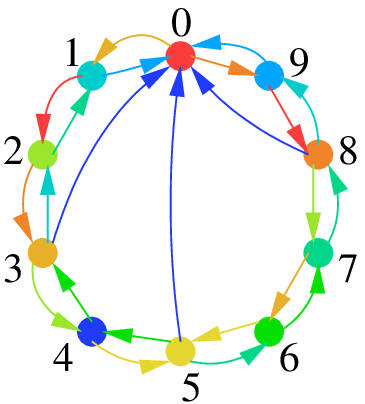}
\hspace{0.5cm}
\includegraphics[scale=0.75]{legend_color.eps}
\hspace{0.5cm}
\includegraphics[scale=0.75]{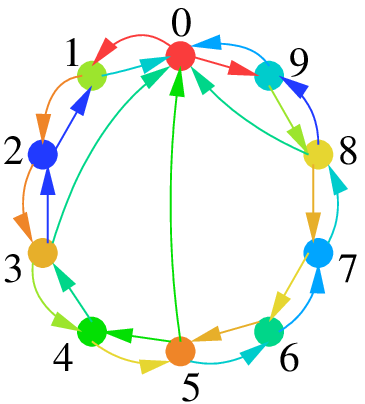}
\caption{\label{fig:vertexedgeranking_Gin1} 
Directed small world network $G_2$. 
Vertex and edge ranking based on Loop Ranking (on the left) or Betweenness Centrality (on the right).}
\end{figure}

Graph $G_2$, shown in figure \ref{fig:vertexedgeranking_Gin1}, represents a
slightly different type of small world network than $G_1$.  The extra
short-cut edges are in this case pointing towards vertex $0$,
increasing its role in dispatching information along the network.  We
also expect the starting vertices $3,5,8$ of the short-cut edges to
play an increased role.

The various ranking results are also reported in
table~\ref{tab:G_vertex}.  All rankings recognize the vertex $0$ as
the more important one.  Loop Ranking and BC 
also capture the special status of the vertices 3, 5 and 8,
even if in a different order.  This is not the case for PageRank, due
to the fact that it does not consider the out-degree of a given vertex
to be very relevant.  A comparison of the Loop Ranking of both types
of graphs shows that the in- and out-bound edges of the vertices are
treated in an identical way, resulting in the same set of Loop Ranking
values for all vertices.  This is not the case for BC 
(even though it does result in the same ordering for both
graphs), as it does not rely on all, but only on the shortest paths.

Figure \ref{fig:vertexedgeranking_Gin1} also includes a schematic 
representation of the edge and vertex ranking according to 
Loop Ranking and BC of graph $G_2$. 
In the case of graph $G_2$, the edges leading away from vertex 0
play a more significant role, while in graph $G_1$ the edges
going to vertex 0 are more important for obvious reasons. 

In conclusion, Loop Ranking reflects the role of vertices and edges
during the dissipation of information along the network.  We have
discussed unweighted small world networks to allow for an easier
comparison between the various ranking methods. Note that Loop Ranking
can naturally be extended to weighted networks, which is a clear
advantage in the analysis of real world networks.

While PageRank is more sensitive to the in-degree of a given vertex,
Loop Ranking treats all paths (passing in either direction through a
given vertex or edge) in an equivalent way.  It produces a slightly
different ordering of importance with respect to BC 
as it takes all paths (with their relative
weight) into account.  Another advantage of the path based rankings we
considered here is that they produce both results for the vertices as
the edges.  Moreover, Loop Ranking can be computed by means of Belief
Propagation, which has a linear time complexity in the system size,
and this is a remarkable practical advantage.  This should also allow
for an easier dynamical analysis of rankings, an aspect which has
already been studied more carefully for BC \cite{bc-dyn}.

A limitation of Loop Ranking is that it is only based on loops. As
such, it can only produce results regarding the vertices and edges
belonging to the 2-core of the graph, i.e., the subgraph for which all
vertices have an in- and out-degree of at least one.  It is reasonable
to assume that the 2-core contains those vertices and edges which are
important to the traffic flow: integration with different schemes
could eventually allow to design ranking methods optimized for
different applications.

This work was supported by EVERGROW, IP 1935 in the complex systems
initiative of the FET directorate of the IST Priority, EU Sixth
Framework. We thank D. Bickson, F. Ricci-Tersenghi and G. Semerjian
for many useful discussions.

\bibliography{paper}

\end{document}